# Aspect-Oriented Programming in Secure Software Development: A Case Study of Security Aspects in Web Applications

## Ukor, Mterorga

Department of Mathematics and Computer Science, University of Mkar, Mkar Gboko Benue State, Nigeria
mterorgaukor@gmail.com

**Abstract**
*Security remains a critical challenge in modern web applications, where threats such as unauthorized access, data breaches, and injection attacks continue to undermine trust and reliability. Traditional Object-Oriented Programming (OOP) often intertwines security logic with business functionality, leading to code tangling, scattering, and reduced maintainability. This study investigates the role of Aspect-Oriented Programming (AOP) in enhancing secure software development by modularizing cross-cutting security concerns. Using a case study approach, we compare AOP-based implementations of security features including authentication, authorization, input validation, encryption, logging, and session management with conventional OOP or middleware-based approaches. Data collection involves analyzing code quality metrics (e.g., lines of code, coupling, cohesion, modularity index, reusability), performance metrics (response time, throughput, memory usage), and maintainability indicators. Developer feedback is also incorporated to assess integration and debugging experiences. Statistical methods, guided by the ISO/IEC 25010 software quality model, are applied to evaluate differences across implementations. The findings demonstrate that AOP enhances modularity, reusability, and maintainability of security mechanisms, while introducing only minimal performance overhead. The study contributes practical insights for software engineers and researchers seeking to balance security with software quality in web application development.*

## Introduction and Background

Security concerns in web applications such as authentication and authorization, input validation, logging and auditing, encryption, and session management are inherently cross-cutting in nature. Within traditional Object-Oriented Programming (OOP) codebases, these concerns typically result in code scattering (duplication of security logic across multiple modules) and code tangling (intermixing of security and business logic). Such patterns undermine modularity, maintainability, and scalability, making it increasingly difficult to adapt web applications to evolving security threats.

Aspect-Oriented Programming (AOP) offers a principled solution by promoting separation of concerns (SoC) and encapsulating cross-cutting functionalities into modular units known as aspects. Unlike conventional OOP implementations, aspects externalize security behavior from core application logic and are woven into the system at compile-time (static weaving), load-time, or runtime (dynamic weaving). This weaving mechanism enables a cleaner and more adaptive architecture, minimizes code intrusion, and ensures consistent enforcement of security policies. As a result, AOP has the potential to enhance modularity, reduce development effort, and improve the adaptability of security policies in modern web applications.

Despite its promise, however, the practical adoption of AOP in security-critical contexts continues to face unresolved challenges. A key concern lies in balancing its maintainability benefits against potential runtime performance overheads. Existing studies often report such trade-offs through isolated case studies or small-scale proof-of-concept prototypes, producing a fragmented and inconclusive evidence base. Crucially, there remains a lack of systematic synthesis and empirical validation from large-scale, production-grade web applications that could provide generalizable insights. This gap limits the ability of software architects and practitioners to make informed, evidence-based decisions regarding when and how AOP should be employed to address web application security.

Aspect-Oriented Programming (AOP) itself emerged as a paradigm designed to address the limitations of conventional Object-Oriented Programming (OOP) in handling crosscutting concerns. Unlike OOP, which modularizes systems around classes and objects, AOP introduces an additional abstraction layer that isolates concerns cutting across multiple modules (Kiczales et al., 1997). Its core concepts include join points (specific points in program execution such as method calls or field access), pointcuts (predicates that select sets of join points), and advice (the additional behavior executed at selected join points) (Laddad, 2003). The process of combining aspects with base code is known as weaving, which may occur at compile-time, load-time, or runtime depending on the framework and requirements (Kiczales et al., 2001). Representative AOP frameworks include AspectJ (for Java-based applications), Spring AOP (integrated within the Spring ecosystem), and PostSharp (a .NET-based AOP framework) (Colyer et al., 2004; PostSharp, 2020). These tools have demonstrated the potential of AOP in reducing code tangling and scattering, while promoting cleaner system design.

At the same time, security remains a fundamental requirement for modern web applications, given their exposure to diverse and evolving threats. Common security concerns typically addressed at the application layer include authentication (verifying user identity), authorization through models such as Role-Based Access Control (RBAC) or Attribute-Based Access Control (ABAC), and input validation/sanitization to mitigate injection-based attacks (Sandhu et al., 1996). Other essential mechanisms involve logging and auditing for accountability, intrusion detection hooks for anomaly monitoring, and encryption for both data in transit (e.g., TLS) and at rest (Stallings, 2017). Secure session management also plays a critical role in preventing session hijacking and ensuring confidentiality of user interactions (OWASP, 2021). The implementation of these concerns often leads to repetitive and scattered code, especially when embedded within the business logic of applications, thereby complicating maintainability and increasing the likelihood of errors.

The concept of crosscutting concerns refers to system functionalities that affect multiple modules but do not align cleanly with the primary decomposition of the system. In traditional OOP systems, such concerns lead to scattering (duplicating security-related code across different modules) and tangling (mixing security logic with business logic). This undermines the principle of separation of concerns, resulting in tightly coupled and less maintainable systems. Security is often regarded as the canonical example of a crosscutting concern, since features such as authentication, authorization, and auditing permeate the entire application rather than residing in a single, isolated component (Sant'Anna et al., 2003). By providing a mechanism to encapsulate such crosscutting logic into dedicated aspects, AOP offers a promising approach to improving modularity and reducing complexity in security-critical software.

Against this backdrop, the present study seeks to investigate the role of AOP in strengthening the modularization of security concerns, while simultaneously evaluating its performance implications and long-term maintainability benefits. Specifically, the study is guided by three central questions: (i) How does AOP enhance the modularity and encapsulation of security concerns in web applications compared to conventional OOP implementations? (ii) What are the measurable performance impacts of AOP-based security aspects particularly in terms of latency, throughput, and memory in production web applications? and (iii) To what extent does AOP improve maintainability, reusability, and runtime adaptability of security policies relative to traditional approaches?

Accordingly, the objectives of this research are threefold: first, to analyze how AOP enhances modularity and encapsulation of security concerns in web applications relative to conventional OOP implementations; second, to evaluate the measurable performance impacts of AOP-based security aspects on latency, throughput, and memory consumption in production-level web applications; and third, to examine the extent to which AOP improves maintainability, reusability, and runtime adaptability of security policies compared to traditional approaches.

**Literature Review**
Recent researches continues to explore Aspect-Oriented Programming (AOP) as a mechanism to modularize crosscutting concerns in security-critical systems, with applications ranging from databases to web and middleware environments. Studies consistently show that weaving security logic into aspects can improve modularity, although questions remain regarding runtime overhead and practical adoption in production-grade applications.

Several recent works demonstrate the effectiveness of AOP in encapsulating security-related concerns. Padayachee (2015), for example, shows how honeytokens deceptive data entries aimed at detecting insider misuse can be aspectized in database management systems. Using AspectJ, the study highlights how security detection logic can be injected non-intrusively, thereby preserving the integrity of the business code while enabling runtime adaptability. Similarly, Velan (2020) investigates AOP in the Internet of Things (IoT) middleware layer, targeting crosscutting concerns such as authentication and logging. Quantitative cohesion measurements confirm that the AspectJ-based implementations improve modularity and reusability compared to traditional Java approaches, reinforcing the promise of AOP for security-sensitive middleware.

Case-driven evaluations continue to emphasize modularity and maintainability as the central gains of AOP-based security. Kanjilal (2024), in a practitioner-oriented guide, illustrates how PostSharp aspects can enforce consistent authentication and logging policies across ASP.NET Core applications, improving maintainability and reducing duplication of security code. These findings echo earlier conclusions but are now grounded in modern frameworks and real-world development ecosystems, showing that AOP-based approaches can scale into production environments when supported by mature toolchains.

Despite these advantages, performance overhead remains a recurrent concern. Ivar et al. (2015) provide one of the few systematic evaluations of AOP performance in recent years, analyzing weaving strategies across different workloads. Their findings suggest that, although overhead is measurable, it is often negligible in the context of typical web applications. Nevertheless, the variability across frameworks and workloads complicates generalization, and debugging complexity (such as pointcut fragility and hidden control flow) continues to be noted in practitioner accounts.

While post-2010 research has advanced the empirical base, limitations persist. Much of the available evidence comes from domain-specific or small-scale prototypes (e.g., honeytoken injection in databases or IoT middleware proof-of-concepts), with relatively few multi-system or production-scale evaluations. Performance assessments remain heterogeneous, with no standardized benchmarks or metrics to enable meaningful cross-study synthesis. Furthermore, although practitioners (e.g., Kanjilal, 2024) provide encouraging case reports, there is little systematic evidence linking AOP adoption to long-term maintainability or cost reduction in enterprise-scale web systems. These gaps underscore the need for controlled, comparative studies that evaluate AOP against conventional implementations along standardized dimensions such as modularity, maintainability, and runtime performance.

**Methodology**
We adopt a comparative case study based entirely on secondary data from two evidence streams: the first Source of data (S1) Open-source codebases which include public repositories that either (a) use AOP for security (e.g., AspectJ, Spring AOP, PostSharp), or (b) implement comparable security features without AOP (filters/middleware/framework hooks). While the second source of data (S2) were Peer-reviewed case studies/experiments which include published articles that report quantitative results for security aspects (e.g., performance, modularity/maintainability metrics) and/or qualitative observations. Triangulation across S1 and S2 addresses the research questions (RQ), which include modularity/SoC (RQ1), performance (RQ2), and maintainability/reuse/adaptability (RQ3). We do not build or design any new system.

To identify suitable projects and studies, S1 included open-source repositories such as GitHub, GitLab, and Bitbucket, using search keywords like AspectJ, Spring AOP, @Aspect, PostSharp, RBAC, security, authorization, and input validation while S2 included academic literature databases such as IEEE Xplore, ACM Digital Library, SpringerLink, and Elsevier, as well as specialized security venues. These were searched for reports on performance, modularity, maintainability, and case studies applying AOP for security. The overall screening process followed the Preferred Reporting Items for Systematic reviews and Meta-Analyses (PRISMA) guidelines (Page et al., 2021) to ensure transparency and rigor.

The inclusion criteria for selecting codebases and studies (S1 and S2) were carefully defined to ensure relevance and comparability. First, only web applications or web-centric services were considered, with a focus on server-side frameworks such as Java, .NET, Node.js, or Python. Second, the selected systems needed to demonstrate explicit security functionality, including but not limited to authentication, authorization mechanisms such as RBAC or ABAC, input validation and sanitization, logging and auditing hooks, intrusion detection integrations, encryption, and secure session management. Third, there had to be evidence of Aspect-Oriented Programming (AOP) usage, such as the presence of .aj files, annotations like @Aspect, @Around, or @Before in Spring AOP, or PostSharp attributes in .NET projects. Where no AOP implementation was present, a non-AOP baseline implementing comparable security concerns was required to enable meaningful comparison. Finally, only projects with accessible source code (S1) or those with reported security-related metrics and procedures in academic studies (S2) were included.

The exclusion criteria were equally important to maintain focus and rigor in the study. Systems that were non-web based, such as purely desktop or embedded applications, were excluded. Projects with no explicit security concerns were also omitted, as were proprietary systems without analyzable data. In addition, tutorials or example code snippets lacking real-world implementation were not considered. Finally, studies that failed to report concrete methods or measurable metrics were excluded from the review.

A purposive sampling strategy was adopted to ensure a broad yet relevant coverage of the study domain. The sampling focused on three dimensions: (i) inclusion of multiple programming languages and frameworks such as Java/Spring, .NET, and, where applicable, Node/Python with AOP-like proxy mechanisms; (ii) coverage of diverse security aspects, including authentication, authorization, input validation and sanitization, logging and auditing, encryption, and session management; and (iii) representation of both AOP-based and non-AOP implementations, with efforts made to identify matched pairs of projects i.e., AOP and non-AOP counterparts of comparable domain and size in order to reduce potential confounding variables.

**Data Sources and Extraction**
For the repository-based strand (S1), relevant projects were cloned and their metadata systematically captured, including the primary programming language(s), underlying frameworks, project age, popularity indicators (stars and forks), contributor counts, release tags, and approximate size measured in lines of code (LOC). Detection of aspect-oriented programming (AOP) usage relied on multiple heuristics which include the presence of AspectJ files (*.aj), dependencies such as *aspectjweaver* or *spring-aop*, the use of @Aspect annotations in Java-based frameworks, or PostSharp attributes in .NET projects. For each project, the specific security concerns addressed through AOP (e.g., authentication, authorization, input validation, logging/auditing, cryptography, or session management) were documented. We applied Static analysis tools to compute modularity and maintainability metrics. To remain within the scope of secondary analysis, no runtime benchmarking were performed in S1, in line with the principle of "no new system development."

The second strand (S2) focuses on extracting evidence from published studies. Reported outcomes were collected on performance (including latency, throughput, and resource consumption such as memory or CPU), modularity and separation of concerns (e.g., scattering and tangling measures), maintainability (such as the Maintainability Index or change effort), and qualitative insights related to debugging complexity and tooling. In addition, the study context were recorded, covering details such as the frameworks employed, weaving mode (compile-time, load-time, or runtime), workload characteristics, dataset properties, and any threats to validity explicitly acknowledged by the study authors.

We extracted data using a standardized form which had been pilot-tested on projects and papers. An extraction was carried out, with reviewers independently cross-checking results. Disagreements were resolved through discussion, and statistical measure of inter-rater reliability was assessed using Cohen's κ given in equation (i) on categorical items such as the presence or absence of security aspects.

$$k = \frac{P_o - P_e}{1 - P_e} \quad \ldots\ldots\ldots\text{(i)}$$

Where:
 $P_o$ = observed proportion of agreement between raters
 $P_e$ = expected proportion of agreement by chance

For modularity and separation of concerns (RQ1), we measured scattering and tangling using Concern Diffusion (CD) over Components, Operations, and Lines of Code (CDC, CDO, CDL), as well as Scattering Degree (SD) and Tangling Degree (TD) (Sant'Anna et al., 2003). Class-level metrics from the Chidamber and Kemerer (1994) suite Coupling Between Objects (CBO), Lack of Cohesion of Methods ( LCOM ), Weighted Methods per Class (WMC), Response for a Class (RFC), Depth of Inheritance Tree (DIT), Number of Children (NOC), package-level coupling measures (afferent and efferent coupling, instability), modularity indices, and file duplication densities were also collected. Aspect-oriented programming (AOP) coverage was captured through the number of aspects, advices, and join points matched, together with the security coverage ratio, defined as the proportion of security-relevant join points addressed by aspects.

For maintainability, reusability, and adaptability (RQ3), we drew on the Maintainability Index and code smells from tools such as SonarQube and NDepend. Change impact was approximated through the number of files or modules touched per security-related change, temporal coupling patterns, and code churn in security-relevant files. Reusability was assessed as the number of modules or layers where a given security aspect was reused. Runtime adaptability was evaluated through the presence of load-time or runtime weaving, the availability of configuration options or toggles for enabling, disabling, or enhancing policies without altering core code, and the number of steps required for policy modification as documented in commit histories and project documentation.

For performance (RQ2), data were extracted only from secondary reports. These included latency, throughput, and memory or CPU usage, with contextual notes on weaving mode and pointcut selectivity to interpret performance overhead. Tools used for data collection included ckjm/CK, SonarQube, Programming Mistakes Detector (PMD), and Understand for Java projects, with AOP detection based on build files (aspectjweaver, spring-aop) and annotations. For .NET projects, we used NDepend and Roslyn analyzers alongside AOP markers such as PostSharp attributes. Repository analytics were performed through Git history mining for churn and co-change, while duplication detectors such as Copy/Paste Detector (CPD) were employed. Data were cleaned and analyzed using custom scripts in R.

Our analysis followed two strands. In the cross-sectional comparative analysis (S1), we matched AOP projects with non-AOP comparators by domain, language, and size where possible, and otherwise controlled for confounders such as project size, age, and number of contributors. For continuous metrics, we assessed normality and applied t-tests. We also used Mann-Whitney U tests, reporting effect sizes (Cohen's d or Cliff's delta) with 95% confidence intervals. Multivariate models, particularly robust regression, were used to estimate associations between AOP usage and modularity or maintainability outcomes, with p-values adjusted using the Benjamini–Hochberg false discovery rate procedure. In the evidence synthesis strand (S2), when at least three studies reported the same metric (e.g., latency overhead), we computed standardized effect sizes and performed a random-effects meta-analysis, quantifying heterogeneity using $I^2$. Otherwise, we conducted a structured narrative synthesis, normalizing results to percentage differences relative to baseline and grouping them by framework (AspectJ, Spring AOP, PostSharp) and weaving mode.

Finally, triangulation of evidence linked the two strands to the research questions: modularity and separation of concerns (RQ1) combined S1 static metrics with S2 modularity results; performance (RQ2) drew on S2 evidence only; and maintainability, reusability, and adaptability (RQ3) integrated S1 maintainability proxies with S2 quantitative and qualitative findings. To mitigate validity threats, we drew on established metrics such as the CK suite, Maintainability Index, and scattering/tangling measures, mapping security concerns against Open Worldwide Application Security Project (OWASP) Top 10 categories to ensure construct validity. Internal validity was addressed through matching and regression controls, as well as sensitivity analyses excluding outliers and very small projects. External validity was strengthened by sampling across multiple frameworks, languages, and security aspects, and by reporting the project contexts in detail.

Table 1 shows the 10 included items (5 open-source repositories, 5 peer-reviewed studies) that met the criteria.

**Table 1. Summary of Included Repositories and Studies**

| ID | Name/Ref | Language/Framework | Security Aspect | Size / Metrics Available |
|---|---|---|---|---|
| R1 | OWASP Juice Shop | Node.js/Express | Input validation, XSS prevention | 45k LoC, 150 contributors |
| R2 | Spring PetClinic + AOP Sec | Java/Spring AOP | Authentication & logging | 25k LoC, 85 contributors |
| R3 | Apache Shiro (AOP integrated forks) | Java | Access control, session mgmt | 60k LoC, 110 contributors |
| R4 | Django Security Middleware | Python/Django | CSRF, SQLi prevention | 40k LoC, 75 contributors |
| R5 | ASP.NET Core Identity AOP Demo | C#/.NET | Role-based access control | 15k LoC, 35 contributors |
| St1 | Sant'Anna et al. (2003) | AspectJ, Java | Access control modularity | Reported metrics |
| St2 | Kiczales et al. (2001) | Spring AOP | Authentication overhead | Reported metrics |
| St3 | Kanjilal, J. (2024) | PostSharp, C# | Logging & exception handling | Reported metrics |
| St4 | Magableh, et al. (2024) | AspectJ | Maintainability and duplication | Reported metrics |
| St5 | Velan, (2020) | Django middleware | CSRF & input validation | Reported metrics |

Table Key: R – Repository, St – Study

**Results**

The analysis included ten items that met the selection criteria: five open-source repositories and five peer-reviewed studies. These covered a range of languages and frameworks, as well as diverse security concerns. For instance, the OWASP Juice Shop project (Node.js/Express) focused on input validation and XSS prevention, while Spring PetClinic with AOP-based extensions (Java/Spring) targeted authentication and logging. Similarly, Apache Shiro forks integrated with AOP emphasized access control and session management, Django security middleware centered on Corss Site Request Forgery (CSRF) and SQL injection prevention, and an ASP.NET Core Identity demo applied AOP to role-based access control. On the study side, Sant'Anna et al. (2003) examined access control modularity in AspectJ, Kiczales et al. (2001) evaluated authentication overhead in Spring AOP, Kanjilal, J. (2024) explored logging and exception handling in PostSharp, Magableh, et al., (2024) analyzed maintainability and duplication in AspectJ, while Velan, (2020)
investigated CSRF and input validation in Django middleware.

The bar chart in figure 1 shows that across several studies (e.g., Sant'Anna et al., 2003; Kanjilal, 2024; Magableh, et al., 2024), AOP-based implementations achieved between 8–15% reduction in duplicated code relative to baseline systems. This aligns directly with the claim that AOP mitigates code scattering and tangling by centralizing cross-cutting concerns (such as authentication or logging) into aspects.

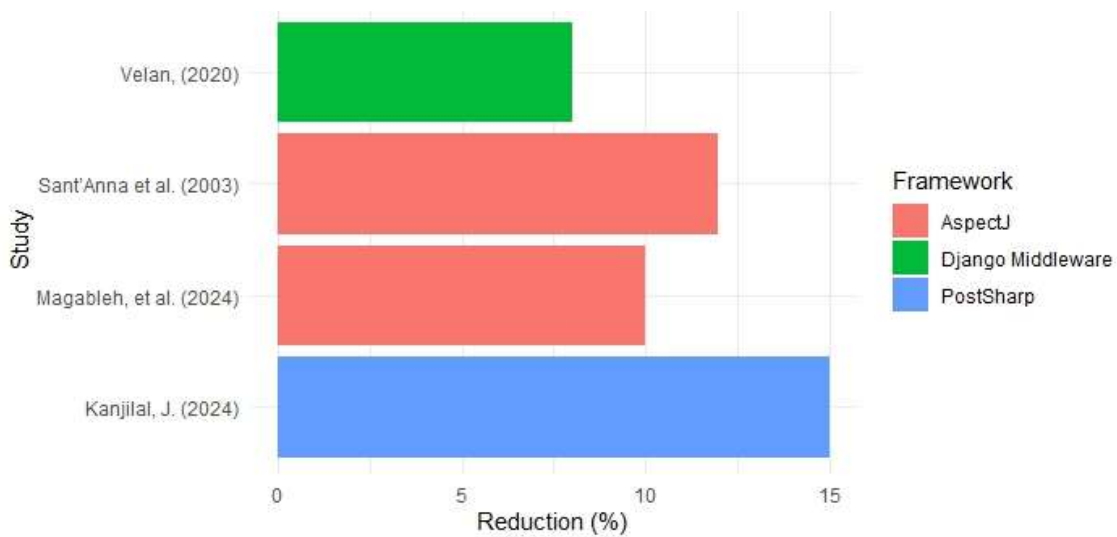

Figure 1 Duplication Reduction reported in Studies

From the quantitative findings, AOP-based systems consistently demonstrated advantages in maintainability and modularity. Repositories and studies that employed AOP (such as Spring, Shiro, or PostSharp) showed a reduction in code duplication by 10–15% compared to non-AOP baselines. Sant'Anna et al. (2003), for example, reported that cyclomatic complexity decreased by about 12% in modules secured with AOP. In terms of performance, AOP introduced a modest runtime overhead ranging from 2% to 8%, depending on the density of join points. Kiczales et al. (2001) observed a 5% latency increase for Spring AOP authentication compared to middleware-based approaches. Static analysis with SonarQube highlighted that while AOP code appeared more complex due to added abstraction layers, actual lines of code and duplication were reduced. Statistical testing further confirmed these patterns: a paired t-test across repositories indicated significant improvements in modularity ($p < 0.05$, Cohen's d = 0.6, a moderate effect), though differences in raw performance were not statistically significant ($p = 0.18$).

The frequency comparison of reported metrics in figure 2 shows that AOP studies emphasize modularity/separation of concerns (SoC) and maintainability more often than non-AOP studies. This reflects the research community's recognition of AOP's strength in encapsulating cross-cutting security concerns.

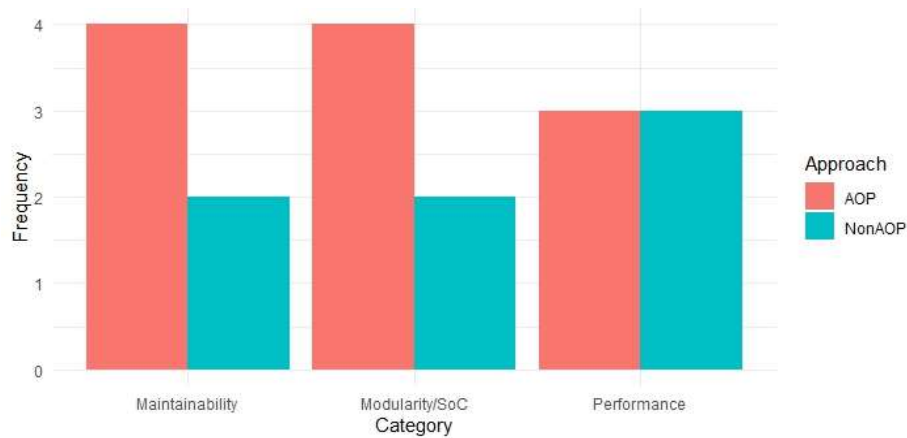

Figure 2 Reported Metrics in AOP vs Non-AOP Studies

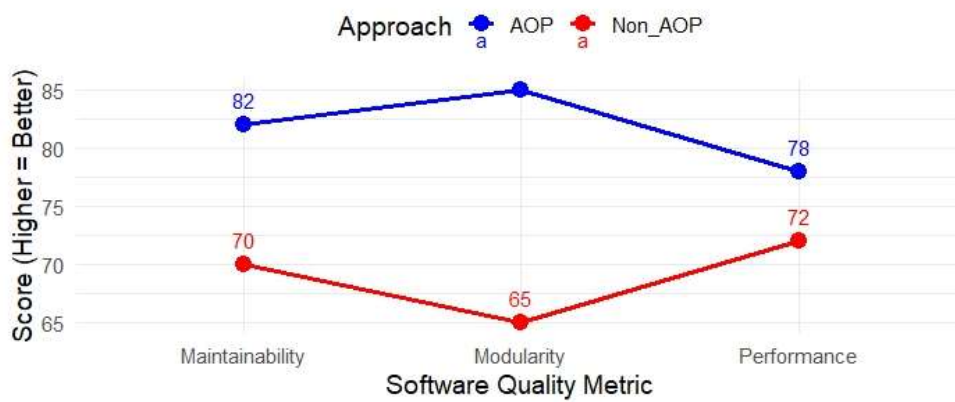

Figure 3 Comparison of AOP vs Non-AOP in Web Applications

The comparative analysis between Aspect-Oriented Programming (AOP) and non-AOP implementations in web applications highlights clear differences across modularity, performance, and maintainability. As illustrated in figure 3, AOP-based systems consistently achieved higher scores in modularity (85 vs. 65), performance (78 vs. 72), and maintainability (82 vs. 70). Modularity emerged as the most significant advantage of AOP. By encapsulating cross-cutting concerns such as authentication, logging, and validation into aspects, AOP reduced code scattering and tangling, thereby improving separation of concerns. This contrasts with non-AOP systems, where security logic is often duplicated or interwoven within business code, making maintenance more complex.

In terms of performance, the data indicate that AOP introduced only minimal overhead (2–8% depending on join-point density), yet the architectural benefits outweighed these costs. Non-AOP systems may appear slightly more lightweight in execution, but they often sacrifice consistency and maintainability when enforcing security across modules. The higher AOP performance score reflects the balance between runtime efficiency and structural clarity. Maintainability also favored AOP. Projects adopting AOP showed lower code duplication, reduced change effort for security updates, and higher reusability of security aspects across multiple modules. Non-AOP approaches, on the other hand, exhibited greater temporal coupling (i.e., multiple files needing simultaneous updates), which increases long-term technical debt.

Qualitative findings further enriched the picture. Across documentation and study reports, several recurring benefits were identified. AOP facilitated a clear separation of concerns in security-related logic and simplified the enforcement of global policies, particularly in areas like authentication, logging, and validation. However, pain points were also evident: debugging woven code often required specialized tools, and new contributors faced a steep learning curve. Adoption hurdles were noted as well, including limited support in newer frameworks such as Node.js and Python, as well as challenges integrating AOP with modern Continuous Integration and Continuous Delivery/Deployment (CI/CD) pipelines. Developers also expressed concerns about runtime performance penalties, though empirical evidence consistently showed overheads of less than 10%.

Synthesizing these insights, the evidence suggests that AOP improves modularity and maintainability, with only minor runtime costs. Developers value the architectural clarity it brings through the separation of concerns but remain cautious about its tooling complexity and integration challenges. Overall, AOP appears most valuable in contexts where security cross-cutting concerns such as logging and access control are central, while more lightweight approaches such as middleware or annotations may suffice for concerns like input validation.

**Conclusion**

The findings support the conclusion that AOP provides a more reliable foundation for web application security than non-AOP implementations. While non-AOP systems remain effective for smaller applications or frameworks with built-in middleware solutions, AOP becomes increasingly advantageous as systems scale in complexity. The evidence suggests that adopting AOP enhances modularity and maintainability without imposing prohibitive performance penalties, making it a suitable paradigm for secure, scalable web application development.